\documentclass[12pt]{JHEP3}
\usepackage{graphicx,amsmath,amssymb}
\usepackage{epsfig,multicol}
\usepackage{epsf}
\input{epsf.sty}

\def\cO{{\cal O}}

\def\m10{M_{10}}

\def\half{\frac{1}{2}}

\def\baray{\begin{eqnarray}}
\def\earay{\end{eqnarray}}

\newcommand{\be}{\begin{equation}}
\newcommand{\ee}{\end{equation}}
\newcommand{\bea}{\begin{eqnarray}}
\newcommand{\eea}{\end{eqnarray}}
\newcommand{\barr}{\begin{array}}
\newcommand{\earr}{\end{array}}

\newcommand{\ba}{\begin{eqnarray}}
\newcommand{\ea}{\end{eqnarray}}

\title{Is Brane Inflation Eternal?}
\author{Xingang Chen$^{1,2}$, Sash Sarangi$^2$, S.-H. Henry Tye$^3$ and Jiajun Xu$^3$ 
\\ \small{\em $^1$Institute for Fundamental Theory \\
Department of Physics, University of Florida,
Gainesville, FL 32611}
\vskip .1cm
\\ \small{\em $^2$Institute of Strings, Cosmology and Astroparticle Physics \\
Physics Department, Columbia University, New York, NY 10027}
\vskip .1cm
\\ \small{\em $^3$Newman Laboratory for Elementary Particle Physics \\
Cornell University, Ithaca, NY 14853} }

\abstract{
In this paper, we show that eternal inflation of the random walk type
is generically absent in the
brane inflationary scenario. Depending on how the brane
inflationary universe originated, eternal inflation of the false
vacuum type is still quite possible. 
Since the inflaton is the position of the $D3$-brane relative to the
$\bar{D}3$-brane
inside the compactified bulk with finite size, its value is bounded.
In DBI inflation, the warped space also restricts the amplitude of the
scalar fluctuation.
These upper bounds impose strong constraints on the 
possibility of eternal inflation. 
We find that eternal inflation due to the random walk of the inflaton
field is absent in both the KKLMMT slow roll scenario and the DBI
scenario.
A more careful analysis for the slow-roll case is also presented using the
Langevin equation, which gives very similar results.
We discuss possible ways to obtain eternal inflation of the random
walk type in brane inflation.
In the multi-throat brane inflationary scenario, the
branes may be generated by quantum tunneling and roll
out the throat. Eternal inflation of the false vacuum type 
inevitably happens in this scenario due to the
tunneling process. Since these scenarios have different cosmological
predictions, more data from the cosmic microwave background radiation
will hopefully
select the specific scenario our universe has gone through.
}

\preprint{hep-th/0608082 \\ CU-TP-1156 \\ UFIFT-HEP-06-11}

 
\begin{document}

\section{Introduction}

	The idea of eternal inflation was introduced more than 20 years ago,
and has been considered as a generic phenomenon common to a wide class
of inflationary models \cite{Steinhardt,Linde:1982,Vilenkin:1983xq,Linde:1986}. 
If eternal inflation is present in the inflationary universe, the inflating part of the 
universe is exponentially larger than the non-inflating part (which includes our 
observable universe) today. This suggests that it is exponentially more likely for 
a random occurrence to take place in the part of the universe that is still inflating.
This will force us to invoke something like the anthropic principle in order to explain 
our very existence. Although a weak form of such an anthropic
principle may be acceptable, 
philosophically at least, the absence/presence of eternal inflation 
is an interesting question. 

There are two types of mechanisms that drive
eternal inflation. The first type involves a meta-stable false
vacuum in the inflaton potential \cite{Vilenkin:1983xq}. The false vacuum
decays exponentially, but at the same time, the volume occupied by it
is expanding exponentially. In most successful inflation models,
the expansion rate is much faster than the decay rate, the false
vacuum never totally disappears and the total volume of it,
once inflation starts, continues to grow exponentially with time. So
there are always bubbles of false vacuum in the entire universe where
inflation never ends. We call this eternal inflation of the false
vacuum type. This is different from the original graceful exit problem
since inflation with enough e-folds happened after tunneling.

The second type of eternal inflation was
proposed in the framework of chaotic inflation \cite{Linde:1993xx}. In this
scenario, quantum fluctuation and ``no hair'' theorem of the de-Sitter
space make eternal inflation possible. The inflaton field is subjected
to Brownian motion due to the superposition of classical motion and
quantum fluctuation. There exists a certain critical field value above 
which the quantum fluctuation dominate over the classical motion. If 
inflation starts above the critical field value, the inflaton field
may fluctuate up the potential quantum mechanically in at least one Hubble
patch. Since the de-Sitter space has event horizon given by the inverse 
Hubble parameter $H^{-1}$, different
Hubble patches evolve independently. The patch
with inflaton field fluctuating up will expand at a
faster rate and produce more Hubble patches, among which there will be
again at least one patch with inflaton field even higher uphill. This leads
to the self-reproduction of the inflationary domains and the process
will repeat eternally. We call this eternal
inflation of the random walk type.

Eternal inflation of the random walk type is natural in
chaotic inflation since it assumes inflation starts with all possible initial
conditions including those initial values well above the critical
value and trans-Planckian. In more general slow-roll inflationary
models, whether such type of eternal inflation can happen depends on
the assumption on the extension of the inflaton potential. For the single
field inflation, the inflaton is responsible for both the
inflation and density perturbations. The density perturbations
$\delta_H$ has to be of order $10^{-5}$ in our observational
horizon with $N_e\sim 55$. If the density
perturbations can reach $\delta_H \sim 1$ when we extend the inflation
back to larger $N_e$, then the quantum
fluctuations of the inflaton will compensate against the classical
rolling downhill and the eternal inflation
of the random walk type will be realized. 

At first sight, eternal inflation appears to be disconnected from
experiments, since it is a statement about universes outside of our
observational horizon. However,
because of the potentially controllable UV origins of string inflation
models such as the brane inflation, the mechanisms that are responsible
for inflation such as the flatness and extension of the potential or
the warping of the background can be concretely constructed and
calculated, rather than
hypothesized. Therefore by fitting various predictions of different
brane inflation models to observations, we can narrow down the
parameter space and eventually identify or rule out specific types of
models. Then the very interesting possibility arises that we can make
concrete statement about the existence/absence of eternal inflation and,
if present, even the nature of the eternal inflation that are directly 
responsible for our own universe. 

In this paper, we consider the possibility of eternal inflation
in brane inflation \cite{Dvali:1998pa}. Brane inflation realizes the inflaton
as a brane rolling inside a stabilized bulk in a string theory flux
compactification, a natural scenario within the brane world framework.
Here the presence or absence of eternal inflation takes a distinctly new
direction. For a field theoretic model with a large field, the
inflaton potential
typically extends to very large values of the inflaton field, 
so that
the critical value are automatically included. In brane inflation,
the brane must sit inside 
the bulk, the inflaton value that measures the brane position is constrained 
by the finite size of the bulk (the finite size of the bulk follows from the finite 
value of the Planck mass). Random walk eternal inflation typically
requires
the critical inflaton value be bigger than the size of the bulk, a
clear impossibility.
In the simplest $D3$-$\bar{D}3$ inflationary scenario
\cite{Kachru:2003sx},
this indeed turns out to be the case. However
for a small field inflation model such as starting off the brane from the
top of an inverted potential, the eternal inflation of the random walk
type is possible. 

In brane inflation, in addition to the usual slow-roll scenario,
there exists another distinctive mechanism
to generate inflation using the speed-limit of the warped space, namely the 
Dirac-Born-Infeld (DBI) inflation
\cite{Silverstein:2003hf,Alishahiha:2004eh,Chen:2004gc,Chen:2005ad}.
The DBI action is dictated by string theory for an open string mode
such as the inflaton in brane inflation.
We again consider the same two types of mechanisms that generate the
eternal inflation. 
Here since the speed limit constraint also applies to
the speed of the inflaton fluctuations in the transverse brane
directions, when we consider the possibility of the random walk
eternal inflation in DBI brane inflation models 
\cite{Alishahiha:2004eh,Shandera:2006ax,Chen:2004gc,Kecskemeti:2006cg},
such a condition can also become important
\cite{Chen:2005ad,Chen:2005fe} in addition to the
condition of the finite bulk size.

To make the discussions more explicit, let us look at several major
classes of brane inflation models.

The typical example of slow-roll brane inflation is the KKLMMT model
\cite{Kachru:2003sx}, where a $D3$ brane is moving towards a $\bar{D}3$ 
brane sitting at the bottom of a warped throat inside the bulk. Flux
compactification
dynamically stabilizes all moduli of the bulk \cite{Giddings:2001yu,Kachru:2003aw}.
The Coulomb attractions between the brane and the antibrane are
red-shifted by the warp factor.  
Since most of the inflation is taking place as the $D3$ brane is
moving down the throat, 
we need to know the properties of the throat only.
The part of the inflaton potential coming from moduli stabilization,
in particular the inflaton mass $m$,  
depends on the details of the flux compactification and is rather
non-trivial to calculate. 
Treating the warp factor $h_{A}$, the total $D3$ charge $N$, the
string scale $\alpha '$,  
the string coupling $g_{s}$ and the inflaton mass $m$ as free
parameters, we can find the critical inflaton value $\phi_{c}$ for  
eternal inflation of the random walk type. Requiring the temperature
fluctuation in the cosmic 
microwave background radiation at $N_{e}=55$ to agree with the COBE data, 
we find that $\beta = m^2 / H^2 < 10^{-8}$, yielding $\phi_c \simeq
0.025M_p$, where $M_{p}$ is the reduced Planck mass. Since the throat
is inside the bulk, the size of this throat must be smaller than the
size of the bulk. To satisfy this constraint while allowing
$\phi>\phi_{c}$ requires $N<1$, which is impossible, since we expect
the $D3$ charge to be large, $N \gg 1$. 
It is possible that the inflaton potential is more complicated than suggested.
However, it is hard to see how this bound can be satisfied even if we
allow substantial fine-tuning.
So we conclude that random walk eternal inflation is ruled out in this
slow roll scenario.

We find it useful to study this scenario more carefully using the
Langevin approach.
It is reassuring that the resulting critical inflaton value $\phi_c
\simeq 0.028M_p$ agrees very 
well with that obtained in the above simple analysis. So this results
in no change in the
conclusion that random walk type eternal inflation is absent in this model.

If the moduli potential which attracts the brane into the throat
becomes very steep, the DBI effect can be triggered. This is the UV DBI
model studied in Ref.~\cite{Silverstein:2003hf,Alishahiha:2004eh}. 
In this model although the brane is rolling slowly, it is
ultra-relativistic because the velocity is restricted by the
speed-limit of the warped space.
As we shall discuss, in
this model, since the running of the density perturbations is very
slow (of order $\cO(1/N_e^2)$), it is impossible for $\delta_H$ to grow
from the observed value $10^{-5}$ at $N_{e}=55$ to 1 anywhere in the throat. 
This model can be smoothly
connected to the slow-roll inflation of the chaotic type once we
extend the brane backward to larger $\phi$ values
\cite{Shandera:2006ax}. Again $\phi_{c}$ cannot be
trans-Planckian in this warped compactification. So we see that 
random walk eternal inflation is ruled out in the UV DBI inflation model. 

In the intermediate regime between the DBI and slow-roll phase, 
where the inflaton is moving slowly but relativistically,
we do not have an analytic analysis. Fortunately, it is not hard to show via 
a numerical analysis that the density perturbation never grows substantially 
above the cosmologically known value of $10^{-5}$ once we impose the 
observational constraint that $\delta_H \sim 10^{-5}$ at $N_{e}=55$. Again, this 
is due to the bound on $\phi$ coming from the size of the bulk.
To conclude, 
an experimental confirmation of such a model would tell us that our
universe does
not directly come from a random walk eternal inflation.

In the case where the mass-squared of the moduli potential becomes
negative, the branes originally trapped inside the throat through some
kind of phase transition will roll out. 
The speed-limit of the IR side
of the warped space can hold these branes on top of the potential for
sufficiently long time to trigger the inflation even if the mass of the
moduli potential is order $H$ or larger. This is the IR model of the
DBI inflation \cite{Chen:2004gc,Chen:2005ad}. 
As we shall discuss, although
the running of the density perturbations in this model is red and
therefore seems to be possible to reach $\delta_H \sim 1$ by
extending the brane deeper into the IR region of the warped space, the
speed-limit constraint also puts non-trivial conditions on 
the size of the scalar
quantum fluctuations. It turns out that such a restriction prevents
the scalar quantum fluctuations from reaching the critical value of the
random walk eternal inflation. 
This conclusion applies to more general DBI inflation models.

It is still possible for eternal inflation of the random walk type to
happen in the bulk, 
especially if a good fraction of the e-folds are generated while the
$D3$ brane is moving
inside the bulk. However, this possibility depends on the detailed
properties of the flux  
compactification. 
%Since there are a vast number of flux
%compactification solutions in 
%IIB theory, it is difficult to address this issue in a definitive way.
For completeness, we give a brief discussion on this possibility,
especially when the $D3$ brane 
is at the top of the inflaton potential in the bulk. We find the
possibility of the eternal
inflation in this scenario depends on  
the likely motion of the $\bar{D}3$ and $D3$ brane at
the initial stage of inflation.

Now we turn to the possibility of eternal inflation of the false vacuum type.
Obviously, we cannot say much about this case without knowing how the brane 
inflationary universe originated. However, we can envision scenarios where
this type of eternal inflation can be compatible with brane inflation.
For example in the IR DBI inflation model, the phase transition that
is responsible for creating the $D3$ branes at the IR end of a throat
may be the flux-antibrane annihilation. This
is caused by the quantum mechanical annihilation between the
antibranes settled 
down in the throat and the fluxes which induces the warped space
\cite{Kachru:2002gs,DeWolfe:2004qx}. Such
a phase transition is of false vacuum eternal inflation type.
Hence an experimental confirmation of this type of brane inflation
model would imply the specific type of the eternal inflation that are
responsible to our universe, namely the false vacuum eternal
inflation.

In this paper, we concentrate on the details of the above several classes
of models. The similar philosophy and procedure can be generalized to other
variations, such as D3-D7 slow-roll brane inflation \cite{Hsu:2003cy} and
multi-brane inflation \cite{Cline:2005ty} etc.

To make the paper more self-contained, we include in the Appendix a brief 
review on the Langevin approaches. It turns out that the 
generalization of the Langevin formulations from the slow roll case to
the DBI case
may be of some interest. To this end, we give a brief discussion on this. 
Besides the non-linear properties of the DBI action, the speed limit on
the inflaton
requires a careful treatment on the noise term to ensure that no superluminal 
phenomenon appears in the ultra-relativistic regime. We present a
simple well-motivated
way to implement the noise term into DBI inflation.
Clearly, this issue requires more analysis.

\section{The absence of eternal inflation in the KKLMMT scenario}

\subsection{The KKLMMT slow-roll scenario} 

In this section, we discuss the possibility of eternal inflation
in the brane inflation models. In a realistic scenario like the KKLMMT
\cite{Kachru:2003sx}, a
$D3$ brane is moving towards a $\bar{D}3$ brane sitting at the bottom of
an AdS throat (the A throat). The metric of the complex manifold takes
the approximate $\textrm{AdS}_5 \times \textrm{X}_5$ form
\cite{Klebanov:2000hb}, and the $\textrm{AdS}_5$ metric is written as
\[
ds^2 = h(r)^2(-dt^2 + a(t)^2 d\vec{x}^2) + h(r)^{-2}dr^2
\]
with the warp factor $h(r) = r/R$ and $R$ is the
curvature radius of the AdS throat. To be specific, let us consider a warped 
deformed $T^{1,1}$ conifold of the Klebanov-Strassler type \cite{Klebanov:2000hb},
\begin{equation}
R^4 \equiv \frac{27}{4}\pi g_s N \alpha'^2 =
\frac{27}{32\pi^2}\frac{N}{T_3} ~,
\end{equation}
where $N=N_{A}$ is the number of the background $D3$ charges, or
equivalently the product of the NS-NS and R-R flux of the throat $A$ (the inflationary throat),
as constructed in \cite{Giddings:2001yu}.
The distance between the two branes is roughly $r$ and the
inflaton field $\phi = \sqrt{T_3}r$ with $T_3$ the $D3$ brane tension. 
In terms of the inflaton potential $V$, the Hubble parameter $H^{2}=V/3 M_{p}^{2}$,
where $M_{p}$ is the reduced Planck mass.
In the case when the brane tension term dominates the potential initially,
it is convenient to write the potential in the following generic form \cite{Firouzjahi:2005dh}
\begin{eqnarray}
V(\phi) &=& V_K + V_A + V_{D\bar{D}} \nonumber \\
&=& \frac{1}{2}\beta H^2\phi^2 + 
\frac{64\pi^2}{27}\frac{\phi_A^4}{N} \left( 1-\frac{1}{N}\frac{\phi_A^4}{\phi^4}
\right) \label{V} ~,
\end{eqnarray}
where the first term $V_K$ receives contribution from K\"ahler potential
and various interactions in the superpotential \cite{Kachru:2003sx} as well as possible
D-terms \cite{Burgess:2003ic}. $V_{A}$ is the tension term that drives inflation and the last term
is the attractive Coulomb-like potential between the $D3$ and $\bar{D}3$ branes. 
$\phi_A$ gives the position of the $\bar{D}3$ brane at the bottom of the throat (note that 
$\phi_A=0$ in the absence of deformation). $\beta$
characterizes the mass of the inflaton field. In general, $\beta
\equiv \beta(\rho, \phi)$ is a function of the K\"ahler modulus $\rho$ and
the location of the brane in the AdS throat. $\beta$ is non-trivial to
determine since it receives many contributions
\cite{Berg:2004sj,Shandera:2004zy,Baumann:2006th}. 
In this paper, we take
the approximation that $\beta$ is more or less constant around any
particular throat and treat $\beta$ as a parameter of the potential. For slow roll
inflation, $\beta$ must be small and detailed analysis shows that $0
\le \beta \le 1/7$ \cite{Firouzjahi:2005dh}.

The potential (\ref{V}) has no false vacuum, so we only need to consider
eternal inflation driven by quantum fluctuation. Generically,
the quantum fluctuation can be modeled as a Gaussian random walk with step size
\begin{equation}\label{dphi}
\delta \phi = \frac{H}{2\pi}
\end{equation}
over the horizon time scale $\Delta t \sim H^{-1}$
\cite{Linde:1993xx}. During the same time, the classical motion of the
inflaton field is
\begin{equation}\label{Dphi}
\Delta \phi = |\dot{\phi} \Delta t| = \frac{|\dot{\phi}|}{H} ~.
\end{equation}
Eternal inflation happens only when 
\begin{equation}\label{dphi_ratio}
\frac{\delta \phi}{\Delta \phi} = \frac{H^2}{2\pi|\dot{\phi}|} > 1 ~.
\end{equation}
It is important to observe that (\ref{dphi_ratio}), when evaluated at
55 e-fold before inflation ends, is the same as the COBE
normalization up to a numerical factor of order 1 \cite{Stewart:1993bc},
\begin{equation}\label{cobe_n}
\delta_H = \frac{2}{5} \left(\frac{H^2}{2\pi|\dot{\phi}|}\right) \Big
|_{N_e = 55} = 1.9 \times 10^{-5} ~.
\end{equation}
Naturally, eternal inflation will never happen if inflation only lasts
for 55 e-fold. In most inflationary models, like chaotic inflation,
the total number of e-fold we can get is usually much larger than
55. It is possible that eternal inflation starts at more than 55 e-fold
before the end of inflation. However, brane inflation in an AdS throat is
more restrictive. As the potential (\ref{V}) is valid only inside
the throat, we have a new constraint in addition to COBE normalization,
\begin{equation}\label{inthroat}
\phi_{A} < \phi_E < \phi_{55} < \phi_i \le  \phi_{\textrm{edge}} ~.
\end{equation}
Here, inflation ends when $\phi$ reaches $\phi_E$, when tachyons appear and the  
$D3-\bar{D}3$ brane annihilation begins.
$\phi_{55}$ corresponds to the position of the $D3$-brane at 55 e-fold
before inflation ends, $\phi_i$ is the initial position where
inflation starts, and $\phi_{\textrm{edge}} \equiv \sqrt{T_3}R$ is the edge of the throat.
It is entirely possible that inflation starts when the $D3$-brane is
outside the throat and we will discuss it later.
%However, without fine-tuning, we expect the $D3$-brane to move
%relatively fast in the bulk and so 
%no eternal inflation takes place there.

We shall see that (\ref{cobe_n}) and (\ref{inthroat}) will
restrict $\beta$ to a tiny fraction of the whole parameter
space in the KKLMMT scenario.

\subsection{Eternal inflation in the slow roll scenario}

The inflaton potential in (\ref{V}) can be expanded to order $O(\beta)$ as,
\begin{equation} \label{V1}
V(\phi)
=\frac{64\pi^2}{27}\frac{\phi_A^4}{N}(1-\frac{1}{N}\frac{\phi_A^4}{\phi^4}
+ \frac{\beta\phi^2}{6M_p^2}) ~.
\end{equation}
The model was studied in \cite{Firouzjahi:2005dh} in detail, and we briefly
summarize their results below. 

The slow roll parameters are:
\begin{eqnarray*}
\eta &=& M_p^2\left|\frac{V''}{V}\right| = \frac{\beta}{3} -
\frac{20M_p^2\phi_A^4}{N\phi^6} ~, \\
\epsilon &=& \frac{M_p^2}{2}\left(\frac{V'}{V}\right)^2 =
\frac{1}{2}\left[ \frac{4M_p\phi_A^4}{N\phi^5} +
\frac{\beta\phi}{3M_p}\right]^2 ~.
\end{eqnarray*}
The end of inflation is given by $\eta = -1$
\begin{equation}\label{phie}
\phi_E^6 = \frac{1}{1+\beta/3} \left(\frac{20\phi_A^4}{N
M_p^4}\right)M_p^6 ~.
\end{equation}
The inflaton field $\phi_N$ is at $N_e$ e-fold before the end of
inflation, and the scalar density perturbation
$\delta_H$ are given by
\begin{eqnarray}
\phi_N^6 &=& \frac{\phi_A^4}{N M_p^4}(24N_e+20) \Omega(\beta) M_p^6 ~,
\label{phiN}\\
\delta_H &=& \frac{2}{45}(24N_e+20)^{5/6}\left(\frac{\phi_A^4}{M_p^4
N}\right)^{1/3} f(\beta) ~, \label{delta_H}\\
\Omega(\beta) &\equiv& \frac{(1+2\beta) e^{2\beta N_e} -
  (1+\beta/3)}{2\beta(N_e+5/6)(1+\beta/3)} ~, \nonumber \\
f(\beta) &\equiv& \frac{1+\beta/3}{1+2\beta} \Omega(\beta)^{5/6}
e^{-2\beta N_e} ~. \nonumber 
\end{eqnarray}
We can use (\ref{delta_H}) to express $\phi_A^4/N$ as a function of $\beta$,
\begin{equation}\label{alpha}
\frac{\phi_A^4}{M_p^4 N}=\left(\frac{45\delta_H}{2}\right)^3(24N_e +
20)^{-5/2} f(\beta)^{-3} ~.
\end{equation}
Given the value of $\beta$, we can impose the COBE
normalization $\delta_H = 1.9 \times 10^{-5}$ at $N_e = 55$ \cite{Bennett:1996ce}, and fix
$\phi_A^4/N$. For the slow roll regime $0 \le \beta \le 1/7$, we get
\begin{equation}
1.2 \times 10^{-18} \le \frac{\phi_A^4}{M_p^4 N} \le 3.5 \times
  10^{-12} ~,
\end{equation}
where the lower bound corresponds to $\beta = 0$ and the upper bound
corresponds to $\beta = 1/7$.

Using (\ref{dphi}) and (\ref{Dphi}), the quantum fluctuation and
classical motion of $\phi$ are given respectively as
\begin{eqnarray}
\label{defindelta}
\delta\phi &=& \frac{4}{9}\sqrt{\frac{\phi_A^4}{N M_p^4}} M_p ~,
\label{dphi_qu} \\
\Delta\phi &=& \frac{4\phi_A^4}{N M_p^4}
\left(\frac{\phi}{M_p}\right)^{-5} M_p + \frac{1}{3}\beta\phi ~.
\label{dphi_cl}
\end{eqnarray}
To compare the magnitude of quantum fluctuation and classical motion,
it is important to notice that $\Delta\phi$ has a minima,
\bea
(\Delta\phi)_{\textrm{min}} = \frac{24}{60^{5/6}} \left
[ \frac{\phi_A^4}{N M_p^4} \beta^5 \right]^{1/6} M_p ~.
\label{dphi_min}
\eea
If for certain values of $\beta$, $\Delta\phi_{\textrm{min}} > \delta\phi$, the classical motion always dominates over the quantum
fluctuation, and eternal inflation will never happen. Using
(\ref{dphi_qu})(\ref{dphi_min}), the range of $\beta$ is determined by
\[
\frac{4}{9}\sqrt{\frac{\phi_A^4}{N M_p^4}} < \frac{24}{60^{5/6}}
\left[ \frac{\phi_A^4}{N M_p^4} \beta^5 \right]^{1/6} ~,
\]
where $\phi_A^4/N$ is also a function of $\beta$ as seen in (\ref{alpha}).
Solving the inequality numerically, we get
\begin{equation}
3.4\times 10^{-8} < \beta \le 1/7 ~.
\end{equation}
Apparently for the slow roll regime $0 \le \beta \le 1/7$, eternal
inflation is possible only when 
\bea
\beta < 10^{-8} ~. 
\label{betacond}
\eea
This indicates 
%that the inflaton mass $m^2 < 10^{-25}M_p^2$, which is 
a large fine tuning, since the moduli stabilization generically gives
$\beta \sim 1$ \cite{Kachru:2003sx}.

For the range $0 \le \beta \le 10^{-8}$, $f(\beta)\approx 1$,
$\Omega(\beta)\approx 1$, the $\beta$ dependence becomes so small that
it is safe to simply set $\beta = 0$. Now, using Eq.(\ref{defindelta}), $\Delta\phi < \delta\phi$ indicates
that eternal inflation happens when
\begin{equation}\label{phic}
\phi > \phi_c = \left( \frac{81\phi_A^4}{N M_{p}^4} \right)^{1/10} M_p
= 0.025 M_p  ~.
\end{equation}
There appears a critical value of the inflaton field $\phi_c = 0.025
M_p$. Inflation will be eternal if it starts with $\phi_i > \phi_c$.

Now we apply the constraint (\ref{inthroat}). To realize eternal
inflation, we need 
\bea
\phi_E < \phi_{55} < \phi_c < \phi_{\textrm{edge}} ~.
\label{phicond}
\eea
Using (\ref{phiN}), we get $\phi_{55} = 3.4 \times 10^{-3} M_p \ll
\phi_c$, so we will surely get enough number of e-fold if inflation
starts with $\phi_c$. However, the constraint $\phi_c < \phi_{\textrm{edge}}$ is non-trivial, since
$\phi_{\textrm{edge}}$ depends on $T_3$ and $N$,
\[
\phi_{\textrm{edge}} = \sqrt{T_3}R = \left( \frac{27N T_3}{32\pi^2}
\right)^{1/4} ~,
\]
where $\phi_c / \phi_{edge} < 1$ requires 
\begin{equation}
N >  \frac{256\pi^5}{27} g_s \left(\frac{\phi_c}{M_p}\right)^4 (\alpha' M_p^2)^2 \approx 10^{-3} g_s (\alpha' M_p^2)^2 ~.
\label{N_lbound}
\end{equation}

There is another constraint we have to take into account. 
Assuming the Planck mass is dominantly contributed by the bulk volume,
as it typically happens to a warped compactification, we have 
\begin{equation} 
M_p^2 =\frac{m_s^8 L^{6}}{\pi g_s^{2}} ~,
\end{equation}
where $g_s$ is the string coupling and $m_s =1/\sqrt{\alpha'}$.
Here $2\pi L$ is roughly the dimension of the bulk.  
Since the throat is part of the bulk, we naturally require
\begin{equation}
L > R ~. 
\label{bulk}
\end{equation}
The requirement (\ref{bulk}) implies an upper bound for the AdS throat charge,
\begin{equation}
N < \frac{4}{27}\left(\frac{g_s}{\pi}\right)^{1/3}
(\alpha'M_p^2)^{2/3} ~. \label{N_ubound}
\end{equation}

It is obvious that the lower bound of $N$ (\ref{N_lbound}) cannot exceed the upper bound (\ref{N_ubound}), so we require
\[
\frac{256\pi^5}{27} g_s \left(\frac{\phi_c}{M_p}\right)^4 (\alpha'
M_p^2)^2 < \frac{4}{27}\left(\frac{g_s}{\pi}\right)^{1/3}
(\alpha'M_p^2)^{2/3} ~,
\]
which leads to a constraint on $\alpha'$
\begin{equation}
\alpha'M_p^2 < \frac{1}{(2\pi)^4\sqrt{2g_s}} \left(\frac{\phi_c}{M_p}\right)^{-3}~. \label{alpha_bound}
\end{equation}

The upper bound of $N$ (\ref{N_ubound}) is maximized if we push $\alpha'$ to the upper bound given in (\ref{alpha_bound}). Thus to make eternal inflation possible, the largest value of $N$ is
\begin{equation}
N < \frac{1}{(2\pi)^3}\frac{4}{27}\left(\frac{\phi_c}{M_p}\right)^{-2}
= 0.95 ~.
\label{Ncond}
\end{equation}

In the KKLMMT scenario, typically $N \gg 1$. This means that if we assume the inflaton potential in the bulk is not very flat and inflation only happens in the throat, the critical initial value $\phi_c$ is impossible  to be consistently realized in the KKLMMT scenario. That is, the random walk eternal inflation is absent.

There are at least 3 possible ways for random walk eternal inflation to be present in this scenario. 
Suppose the potential has additional contributions (see
e.g. \cite{Baumann:2006th}). That may change the above analysis if the potential is changed substantially. The presence of extra terms in the
inflaton potential depends on the details of the flux
compactification. It will be interesting to see if their presence
relaxes the constraints enough to allow eternal inflation.

Another possibility is by allowing the brane to start in the bulk
instead of only from the UV side of the throat; this may allow us to
evade the constraint (\ref{Ncond}).  The antibrane in the end of the
throat provides a red-shifted Coulomb attraction source and the
potential is similar to the first two terms in the (\ref{V1}). Assume
one can tune away the potential from the moduli stabilization (in the
same sense as requiring an extremely small $\beta$), the relaxed
constraint on $\phi_c$ is
\[
\phi_c < L\sqrt{T_3} ~,
\]
which implies 
\begin{equation}
L > (2\pi)^3 g_s \alpha'^2 \phi_c^2 ~.
\label{Lcond}
\end{equation}
The condition (\ref{bulk}) still applies here so we still have the
same upper bound of $N$ as in (\ref{N_ubound}).  It is obvious that
(\ref{Lcond}) and (\ref{N_ubound}) are not contradictory and by
choosing appropriate values of $L$ and $\alpha'$, we can have $N \gg
1$.  So in the case when we get a good fraction of e-folds from the
bulk, eternal inflation is possible when $\beta < 10^{-8}$ and
inflation starts with $\phi_i > \phi_c$. 
We discuss more on the possibility of the eternal inflation in the
bulk in Sec.~\ref{SecTop}.

Note that the bound (\ref{bulk}) applies to a throat whose length is comparable to its base width.
If the throat is a long narrow throat (that is, its width is much smaller than its length), then the bound is relaxed. Orbifolding a deformed conifold  can be a way to achieve this. Even if this bound is relaxed, we note that  the bound (\ref{betacond}) is still very tight.

\section{Langevin analysis for the slow-roll scenario}

Here we shall evaluate the critical initial value $\phi_{c}$ via the Langevin approach.
A brief review of the background
can be found in the Appendix. Since brane inflation includes the DBI action automatically, 
the Appendix presents a brief discussion starting with a Langevin formalism for the DBI case.
In this section, we only need the more standard and relatively well
known formulation in slow-roll case. 

We want to explore how eternal inflation will affect the way we
extract predictions from theoretical models.  The ``no hair''
theorem of de-Sitter Space allows inflaton
field in different Hubble patches to evolve independently. If in all
of the Hubble patches, the inflaton field follows more or less the
same path rolling down the potential, then we can simply use our
prediction in one Hubble patch for all of the other ones. But if there
is some patch in the universe that is eternally inflating, then that
patch actually dominates the universe, and all the patches where
inflation have ended are exponentially small compared to that one. In
this case, we need to look at the statistical average of physical 
observables over all the Hubble patches. In the following discussion,
we shall choose the Hubble constant as an example. 

To perform the average, we need to define the statistical measure. The choice of
measure is quite non-trivial and does change the result of
prediction, discussions can be found in
e.g. \cite{Tegmark:2004qd,Gratton:2005bi,Winitzki:2005ya}. 
In our calculation, we shall use the
idea from the Fokker-Planck approach \cite{Linde:1993xx} and define
the measure to be the physical volume of the Hubble patch. 

The average over the physical volume is defined as
\begin{eqnarray}
\langle H(t) \rangle_p &=& \frac{\langle H(t) e^{3N(t)}
  \rangle}{\langle e^{3N(t)} \rangle} ~, \label{H_averp}\\
N(t) &=& \int_0^{t} dt' H(t') ~. \nonumber
\end{eqnarray}
We consider the physical volume to be a good measure to
characterize eternal inflation, because the Hubble patch that is
eternally inflating will have exponentially larger physical volume and
take the largest weight in the average at late
times. Thus the average $\langle H(t)\rangle_p$ will be significantly
changed by quantum fluctuation if eternal inflation happens.

The motion of inflaton field including quantum fluctuation, modeled as 
a random walk, can be described using the Langevin
equation \cite{Linde:1993xx} below,
\begin{eqnarray*}
3H\dot{\phi}+\frac{dV}{d\phi} &=& 3\frac{H^{5/2}}{2\pi}n(t) ~, \\
V(\phi) &=& \frac{64\pi^2}{27}\alpha \left(1-\frac{\alpha}{\phi^4}
\right) ~,
\end{eqnarray*}
where $\alpha \equiv \phi_A^4/N$, and the Langevin force $n(t)$ is a Gaussian stochastic
variable normalized as
\begin{eqnarray}
\langle n(t) \rangle &=& 0  ~, \label{n_aver}\\
\langle n(t_1)n(t_2) \rangle &=& \delta(t_1-t_2) ~, \label{n_std}
\end{eqnarray}
with all higher cumulants being zero.
Here we only consider the case $\beta = 0$ for simplicity.

The equation of motion for the inflaton field is derived from the
Langevin equation as
\begin{equation}
\dot{\phi}+b\phi^{-5} = d M_p^{3/2} n(t)  \label{langevin}
\end{equation}
with the two parameters defined by
\begin{eqnarray*}
b &\equiv&  \frac{32\pi}{9} \left(\frac{\phi_A^4}{N
M_p^4}\right)^{3/2} M_p^7 ~, \\
d &\equiv& \frac{8\sqrt{2\pi}}{27} \left(\frac{\phi_A^4}{N
M_p^4}\right)^{3/4} ~.
\end{eqnarray*}
In the stochastic equation of motion (\ref{langevin}), if we set
$d = 0$, we will recover the usual slow roll equation of motion
and the inflaton field will follow a classical path
$\phi_{cl}(t)$. Since $d \sim 10^{-13}$, we can expand the
solution $\phi(t)$ around $\phi_{cl}(t)$. To order $O(d^2)$, we have
\[
\phi(t) = \phi_{cl}(t)+ d\phi_1(t) + d^2\phi_2(t)+O(d^3) ~.
\]
Substitute the expansion into (\ref{langevin}), we get three coupled
equations by matching the terms of the same order in $d$
\begin{eqnarray*}
\dot{\phi_{cl}} &=& -b \phi_{cl}^{-5} ~, \\
\dot{\phi_1} &=& \frac{5b\phi_1}{\phi_{cl}^6} + n(t) M_p^{3/2} ~, \\
\dot{\phi_2} &=& \frac{5b\phi_2}{\phi_{cl}^6} +
\frac{10b\phi_1^2}{\phi_{cl}^7}  ~.
\end{eqnarray*}
These three equations can be solved analytically and we get the
solution to (\ref{langevin}) as
\begin{eqnarray}
\phi(t) &=& \phi_{cl}(t)+ d \phi_{cl}^{-5}(t)\xi(t) + 
d^2 \phi_{cl}^{-5}(t)\int_0^t dt'
\frac{10b\xi^2(t')}{\phi_{cl}^{12}(t')} ~, \label{phi_t}\\
\phi_{cl}(t) &=& (\phi_0^6-6bt)^{1/6} ~.
\end{eqnarray}
Here $\xi(t)$ is a new stochastic variable defined as,
\[
\xi(t) \equiv \int_0^t dt' n(t') (\phi_0^6-6bt')^{5/6}M_P^{3/2} ~.
\]
The normalization of $\xi(t)$ follows that of $n(t)$ in
(\ref{n_aver}) and (\ref{n_std}),
\begin{eqnarray}
\langle \xi(t) \rangle &=& 0 ~, 
\label{xi_aver}\\
\langle \xi(t_1)\xi(t_2) \rangle &=& \frac{M_p^3}{16b}
\left[\phi_0^{16}-(\phi_0^6 -6b \min[t_1,t_2])^{8/3}\right] ~.
\label{xi_std}
\end{eqnarray}

The Hubble parameter is
\begin{eqnarray}
H(t) &=& H_{cl}(t) + \frac{8\pi}{9 M_p} \sqrt{\alpha} \left[\frac{2d\alpha\xi(t)}{\phi_{cl}^{10}(t)}
         +\frac{3d^2\alpha\xi^2(t)}{\phi_{cl}^{16}(t)}
         +\frac{20d^2\alpha b}{\phi_{cl}^{10}(t)} \int_0^t dt'
         \frac{\xi^2(t')}{\phi_{cl}^{12}(t')} \right]  ~, \label{Ht} \\
H_{cl}(t) &\equiv& \frac{8\pi}{9 M_p} \sqrt{\alpha} \left[ 1 -
         \frac{\alpha}{2\phi_{cl}^4(t)} \right] ~. \label{H_cl}
\end{eqnarray}
Here $H_{cl}(t)$ is Hubble parameter if the inflaton field follows the classical motion.

So far, we have derived the evolution of $H$ in one single
Hubble patch. To calculate $\langle H(t) \rangle_p$ as defined in
(\ref{H_averp}), we shall use the technique developed in
\cite{Gratton:2005bi}. Define a generating functional
\begin{eqnarray*}
W_t[\mu] &=& \ln \langle e^{M_t[\mu]} \rangle ~, \\
M_t[\mu] &=& \int_o^t dt' \mu(t') H(t') ~.
\end{eqnarray*}
$\langle H(t) \rangle_p$ can be evaluated by functionally
differentiating $W_{t}[\mu]$ with respect to $\mu$, and setting $\mu =
3$ later. To order $O(d^2)$ we get 
\begin{eqnarray}
\langle H(t) \rangle_p &=& \frac{\delta W_{t}[\mu]}{\delta \mu(t)}
\Big|_{\mu(t)=3} = \langle H(t) \rangle + 3 \int_0^{t}
dt' \langle\langle H(t)H(t') \rangle\rangle ~, \label{H_aver_p} \\
\langle\langle H(t)H(t') \rangle\rangle &\equiv& \langle H(t)H(t')
\rangle - \langle H(t) \rangle \langle H(t') \rangle ~. \nonumber
\end{eqnarray}

Using (\ref{xi_aver}), (\ref{xi_std}) and  (\ref{Ht}), we can further get
\begin{eqnarray}
\langle H(t) \rangle = H_{cl}(t) &+& \frac{8\pi d^2 \alpha^{3/2}
  M_p^2}{9b} \left[ \frac{19}{48}\frac{\phi_0^{16}}{\phi_{cl}^{16}(t)}
  - \frac{1}{3}\frac{\phi_0^{10}}{\phi_{cl}^{10}(t)} - \frac{1}{16}
\right] ~, \label{1st_cumulant}\\
3 \int_0^{t} dt' \langle\langle H(t)H(t') \rangle\rangle &=& 
\frac{16\pi^2d^2\alpha^3  M_p}{27b} \frac{1}{\phi_{cl}^{10}(t)} 
\int_0^t dt' \left
  [ \frac{\phi_0^{16}}{\phi_{cl}^{10}(t')}-\phi_{cl}^6(t') \right] ~. 
\label{2nd_cumulant}
\end{eqnarray}

To see how quantum fluctuation changes the expectation value of $\langle H(t) \rangle_p$, we look at two limiting cases, $t \ll b^{-1} \sim H^{-1}$ and $t \to T$, where
$T$ is the time when inflation is supposed to end following the classical motion of the branes. $T$ is defined by (\ref{phie})
\[
\phi_{cl}(T) = \phi_E = \left(\frac{20\phi_A^4}{N
M_p^4}\right)^{1/6}M_p ~.
\]
In the case $t << b^{-1}$, the leading order behavior of $\langle H(t) \rangle_p$ in terms of $t$ is
\begin{equation}
\langle H(t) \rangle_p = \langle H(t=0) \rangle_p +
\frac{16\pi}{9M_p}\frac{\alpha^{3/2}}{\phi_0^{10}}(bt) + \frac{8\pi
d^2 M_p^2}{3b}\frac{\alpha^{3/2}}{\phi_0^6}(bt) ~,
\end{equation}
where the second term comes from expanding the classical motion $H_{cl}(t)$ and the last term is from the quantum correction (\ref{1st_cumulant}). It is interesting to notice that the contribution from (\ref{2nd_cumulant}) is of order $(bt)^2$, so it is negligible when $t \ll b^{-1}$.

Requiring that the classical motion is not significantly altered by quantum noise, we need to impose
\[ 
\frac{16\pi}{9M_p}\frac{\alpha^{3/2}}{\phi_0^{10}}(bt) > \frac{8\pi d^2 M_p^2}{3b}\frac{\alpha^{3/2}}{\phi_0^6}(bt),
\]
and this gives the bound
\begin{equation}
\phi_0 < (32\pi)^{1/8}M_p = 1.77M_p ~, \label{bound_i}
\end{equation}
which is much looser than the previous estimate (\ref{phic}).

Now if we look at the late time behavior of $\langle H(t) \rangle_p$ and take the limit $t \to T$, we get
\begin{eqnarray}\label{HT}
\langle H(T) \rangle_p &=& \frac{8\pi}{9 M_p} \sqrt{\alpha} \Big [ \left( 1 - \frac{\alpha}{2\phi_E^4} \right) + \frac{d^2 \alpha M_p^3}{b} \left( \frac{19}{48} \frac{\phi_0^{16}}{\phi_E^{16}} - \frac{1}{3}\frac{\phi_0^{10}}{\phi_E^{10}} - \frac{1}{16} \right) \nonumber \\
& & + \frac{2\pi d^2\alpha^{5/2}\phi_E^2 M_p^2}{3b^2}
\left( \frac{1}{4}\frac{\phi_0^{16}}{\phi_E^{16}} - 
\frac{1}{3}\frac{\phi_0^{12}}{\phi_E^{12}} + \frac{1}{12}\right)\Big] ~.
\end{eqnarray}
The leading order effects of the quantum diffusion is
\[
\frac{19d^2 \alpha M_p^3}{48b} \frac{\phi_0^{16}}{\phi_E^{16}} + 
 \frac{\pi d^2\alpha^{5/2}\phi_E^2 M_p^2}{6b^2}\frac{\phi_0^{16}}{\phi_E^{16}}
= \frac{19 d^2 \alpha M_p^3}{48b} \frac{\phi_0^{16}}{\phi_E^{16}}
 \left[ 1 + \frac{45}{19} \frac{\alpha}{\phi_E^4} \right] ~.
\]
The quantum fluctuation is comparable to classical motion when
\[
1 - \frac{\alpha}{2\phi_E^4} = \frac{19d^2 \alpha M_p^3}{48b} 
\frac{\phi_0^{16}}{\phi_E^{16}} \left[ 1 + \frac{45}{19}
\frac{\alpha}{\phi_E^4} \right] ~,
\] 
which gives the critical initial value
\begin{equation}
\phi_c = 2.1 \left(\frac{\alpha}{M_p^4}\right)^{5/48} M_p = 0.028 M_p
~,
\label{bound_ii}
\end{equation}
which is in very good agreement with the earlier estimate of $\phi_{c}$ (\ref{phic}).

The two bounds (\ref{bound_i}) and (\ref{bound_ii}) can be understood as following. 
If inflation starts at $\phi_0 > 0.028 M_p$, quantum
fluctuation starts to qualitatively change the path of the inflaton
field and we can no longer trust our perturbative
treatment of the Langevin equation. But the effect of the quantum fluctuation does not show up at early time during inflation, as from (\ref{bound_ii}), this requires an even high initial value $\phi_0 = 1.77M_p \gg 0.028M_p$. During time $t \ll H^{-1}$, the motion of the brane is still mostly classical. But as inflation goes on, quantum fluctuation results in different e-folds in different Hubble
patches, and the physical volume average of
observables will be changed more and more, because different Hubble
patches are less and less equally weighed. The Hubble patch
with the largest number of e-fold will occupy the largest physical
volume and dominate the statistical average. All of these are
strong indications that the inflaton field is in a state dominated by
quantum fluctuation, and inflation may go on forever in some Hubble
patch once it starts. 

However, as shown in the earlier section, The above $\phi_{c}$
violates the size of the bulk and so cannot be realized in the KKLMMT
scenario. This rules out eternal brane inflation of the random walk
type in this slow roll scenario.

\section{The absence of eternal inflation in DBI inflation}

In the previous sections we studied the possibility of the eternal
brane inflation of the random walk type in the KKLMMT 
slow-roll scenario.
In this section we study that in the DBI inflation scenario
\cite{Silverstein:2003hf,Alishahiha:2004eh,Chen:2004gc,Chen:2005ad}. 
We shall conclude that eternal brane inflation of the 
random walk type is again absent in the DBI inflation scenario.

\subsection{The UV model}

We consider the case where the inflaton mass gets large, typically
when $\beta \gg 1$. The brane is moving relativistically toward to the IR
end of the AdS throat, even at ultra-relativistic speed. For such a fast roll inflation,
string theory dictates that we need to include higher powers of the
time derivative of $\phi$, in the form of the DBI action \cite{Silverstein:2003hf}
\[
S = -\int d^4x a^3(t)
\left[T(\phi)\sqrt{1-\dot{\phi}^2/T(\phi)} +  V(\phi) - T(\phi)
\right] ~,
\]
where $h(\phi)$ is the warp factor and $T= T_{3}h^4(\phi)$ is the warped tension. 
It is useful to introduce the parameter $\lambda$ via 
$T=T_{3}h^{4}(\phi)=\phi^{4}/\lambda$.

The effective inflaton potential takes the form \cite{Shandera:2006ax}
\begin{equation}\label{V_dbi}
V(\phi) = \frac{1}{2}\beta H^2\phi^2 +
\frac{64\pi^2}{27}\frac{\phi_A^4}{N}
\left(1-\frac{1}{N}\frac{\phi_A^4}{\phi^4}\frac{(\gamma+1)^2}{4\gamma}\right)
\end{equation}
with $\gamma$ the Lorentz factor given by
\begin{equation}\label{gamma}
\gamma^{-1} = \sqrt{1-\frac{\dot{\phi}^2}{T_3 h^4(\phi)}} ~.
\end{equation}
In the non-relativistic limit, $\gamma = 1$, the potential
(\ref{V_dbi}) is the same as the potential (\ref{V}) for the slow roll
inflation. When $\beta \gg 1$, the motion of the brane becomes
ultra-relativistic. The contribution from the Coulomb
potential can be neglected and only the mass term is kept. 

Even if $\beta$ may get very large, making the inflaton potential
steep and causing the brane to move ultra-relativistically
inside the throat, it is still possible to get enough number of e-folds because
while the mobile $D3$ brane approaches the $\bar{D}3$ brane at the
bottom of the throat, $\dot{\phi}$ is bounded from above as can be
seen from the expression of $\gamma$  and the rapid decrease of the wrap factor $h(\phi) \simeq
\phi/\phi_{edge}$ \cite{Silverstein:2003hf}. Together, the Lorentz factor and the large 
warping of the AdS throat acts like a brake on $\dot{\phi}$ and a large number of e-folds
can be completed inside the throat when $\gamma \gg 1$.

Note that
\ba
\frac{\ddot{a}}{a} &=& H^2(1-\epsilon) ~, \\
 \epsilon
&=&\frac{2M_p^2}{\gamma}\left(\frac{H^{\prime}(\phi)}{H(\phi)}\right)^2 ~,
\ea
where $\epsilon$ reduces to the usual slow-roll parameter in the slow-roll approximation.
For inflation to occur $0 \le \epsilon \ll 1$, so this is a good expansion parameter for observational quantities measured $N_{e}$ e-folds back from the end of inflation. By definition, inflation ends when 
$\epsilon=1$. 

The scalar density perturbation is given by
\begin{equation}
\delta_H = \frac{H^2}{5\pi|\dot{\phi}|} \Big |_{k=a\gamma H}
\label{dH_dbi} ~.
\end{equation}
We must evaluate $\delta_H$ at $k = a\gamma H$. 
In terms of a general Lagrangian, this is the effect of
a small sound speed $c_s$
\cite{Garriga:1999vw,Alishahiha:2004eh,Chen:2006nt},
which in this case equals to $1/\gamma$. The
fluctuation mode is not traveling at the speed of light, but instead
at the sound speed. The mode $k$ exits the horizon and freezes
when $kc_s = aH$.

The equations of motion for the inflaton field are \cite{Alishahiha:2004eh,Shandera:2006ax},
\begin{eqnarray}
V(\phi) &=& 3M_p^2H^2(\phi) - T_3h^4(\phi)(\gamma(\phi) - 1) ~, \label{eom_V}\\
\gamma(\phi) &=& \sqrt{1 + 4M_p^4T_3^{-1}h^{-4}(\phi)H'^2(\phi)} ~, 
\label{eom_gamma} \\
\dot{\phi} &=& \frac{-2M_p^2H'(\phi)}{\gamma(\phi)} ~. \label{eom_phidot}
\end{eqnarray}

For the ultra-relativistic case, we keep only the mass term in the
potential (\ref{V_dbi}). Define $m^2 = \beta H^2$, we have
\[
V(\phi) = \frac{1}{2}m^2\phi^2 ~.
\]
The equations of motion can be solved analytically, the leading order
is
\begin{eqnarray}
H(\phi) &=& \frac{\hat{m}}{M_p} \frac{\phi}{\sqrt{6}} ~,\label{H_dbi}\\
\dot{\phi} &=& -\phi^2 \sqrt{\frac{32\pi^2}{27N}} `,\label{phidot_dbi}
\end{eqnarray}
where
\[
\hat{m} \equiv
\left(\sqrt{\frac{64\pi^2}{81N}}+\sqrt{\frac{64\pi^2}{81N}+\frac{m^2}{M_p^2}}\right)M_p
~.
\]
Using (\ref{dH_dbi}), (\ref{H_dbi}) and (\ref{phidot_dbi}), we see
that the scalar density perturbation in leading order 
is a constant independent of $\phi$,
\begin{equation}
\delta_H = \frac{\sqrt{3N}}{40\sqrt{2}\pi^2} \frac{\hat{m}^2}{M_p^2} ~.
\end{equation}
This means if we impose COBE normalization at $N_e=55$, we'll have
the same normalization at any number of e-folds. The fact that
$\delta_H$ has no $\phi$ dependence is largely due to the
ultra-relativistic motion of the brane. In the expression
(\ref{eom_phidot}) for
$\dot{\phi}$, there is the Lorentz factor $\gamma(\phi)$ in addition
to the usual term $H'(\phi)$ for the external force. In the slow roll
scenario, the $\gamma$ factor would not be there and there will be $\phi$
dependence in $\delta_H$. Recall the argument
(\ref{dphi_ratio}), we see that for ultra-relativistic fast roll
inflation, $\delta \phi \ll \Delta\phi$ all the way through
the throat no matter where inflation starts. For UV model with
$\gamma \gg 1$, eternal inflation will not happen.  

\subsection{The intermediate case}

It has been shown in Ref.\cite{Shandera:2006ax} that there are three interesting regions in the parameter space of DBI inflation (i.e., the power spectrum is red-tilt.). The effective potential of DBI inflation takes the form (\ref{V})
\begin{eqnarray}
V(\phi) &=& V_K + V_A + V_{D\bar{D}} \nonumber \\
&=& \frac{1}{2}\beta H^2\phi^2 + 
\frac{64\pi^2}{27}\frac{\phi_A^4}{N} \left( 1-\frac{1}{N}\frac{\phi_A^4}{\phi^4}
\right) ~, \nonumber
\end{eqnarray}
where $H^2 \sim V_A$ for slow roll case. For DBI inflation, we parameterize the inflaton mass the same way even if $H$ may not necessarily be determined by $V_A$.

The three regions are essentially characterized by the mass of the inflaton field :
\begin{itemize}
\item The slow roll scenario when $\beta \ll 1$, the constant term $V_A$ dominates the potential. The inflaton is always non-relativistic; that is $\gamma \simeq 1$. 
\item the ultra-relativistic scenario when $\beta \gg 1$, the mass term dominates the potential. Here $\gamma \gg 1$ so the speed limit effect is significant. It was pointed out that the non-Gaussianity in the cosmic microwave background radiation can be substantial \cite{Alishahiha:2004eh}.
\item the intermediate scenario when $\beta \sim 1$, the mass term and the constant piece are comparable and they interplay with each other. Generically the brane starts with $\gamma \simeq 1$ and its motion becomes ultra-relativistic at the end of inflation. Tensor mode in the cosmic microwave background radiation can be substantial.
\end{itemize}

So far, we have shown that eternal inflation is generically absent in
the first two scenarios. Now we like to consider the intermediate
scenario. First, we note that the intermediate case can be viewed as a
combination of the slow-roll case and the relativistic case. Inflation
starts in slow-roll type and ends in the ultra-relativistic type. At
$N_e=55$, we know that $\delta_H\sim 10^{-5}$. If we try to look at
the region $N_e > 55$, we will find the transition to the 
slow-roll inflation of the chaotic
type. We know that the eternal inflation would happen if $\phi>M_p
\sqrt{M_p/m}$, but such a large $\phi$ exceeds the bulk size.
%we generically find slow roll inflation with $\beta \sim 1$, we know
%from (\ref{betacond}) that eternal inflation is absent in this case. 
If we look at $N_e < 55$, then inflation is
generically of the relativistic type, and from the discussion above,
we know that eternal inflation is again generically impossible. So we
conclude that in the intermediate case, we have a certain ``no go''
scenario that generically forbids eternal inflation. This result is
confirmed by a numerical analysis.

\subsection{The IR model}
\label{SecIR}

We have seen in the previous studies that, in order to have the random
walk type eternal inflation, two requirements are needed. Firstly, the
spectral index has to be red ($n_s-1<0$), so that in the far past,
the condition (\ref{dphi_ratio}) or equivalently 
$\delta_{H} \gtrsim 1$ may be
satisfied. Secondly, the potential extension has to be long enough to
reach this critical 
value. In the KKLMMT type slow-roll brane inflation, the
second requirement imposes a strong constraint on the parameter space,
while in the UV model of the DBI inflation, the first requirement is
not
satisfied in the AdS throat ($n_s-1$ starts at order $\cO(1/N_e^2)$). 
In the IR model of the DBI
inflation, the brane is moving out from
the IR side of the warped space. During most of the inflationary
epoch, the Hubble parameter is approximately a constant and a de
Sitter phase of inflation is triggered.
The field theory calculations \cite{Chen:2004gc,Chen:2005ad}
show that the spectral index is red and
the inflaton can be extended back in time to the deep IR
region of the AdS throat. So naively the eternal inflation of random
walk type is possible. But interestingly as we will see, by
considering 
the validity of the field theory analyses, 
a rather different mechanism \cite{Chen:2005ad,Chen:2005fe} kicks in
before the condition (\ref{dphi_ratio}) can be satisfied.

When calculating the quantum fluctuations in the expanding background, 
for the field theory to work, we require that the Hubble energy be
smaller than the (red-shifted) string scale,
\bea
\gamma H \lesssim T_3^{1/4} h ~,
\label{Hcond}
\eea
where $H$ is the Hubble constant, $T_3$ is the brane tension and $h =
\frac{\phi}{\lambda^{1/4} T_3^{1/4}}$
is the warp factor.
It is easiest to get Eq.~(\ref{Hcond}) on the moving frame on the brane,
where the Hubble expansion is faster by a factor of
$\gamma$ because of the relativistic time dilation. A closely related
consequence is the reduction of the horizon size by a factor of
$\gamma^{-1}$. 

The evolution of the brane position (inflaton) is given by 
\bea
\phi \approx -\frac{\sqrt{\lambda}}{t} ~,
\eea
where $t$ runs from $-\infty$. The inflation ends when $t\sim -1/H$. The
number of inflationary e-folds to the end of the inflation is
\bea
N_e \approx -H t ~.
\eea
During inflation the inflaton travels ultra-relativistically,
\bea
\gamma \approx \frac{\beta N_e}{3} ~,
\eea
where $\beta$ again parameterizes the steepness of the potential $V=
V_0 - \half \beta H^2 \phi^2$.
Using these field theory results of the IR DBI inflation,
we translate (\ref{Hcond}) into
\bea
N_e \lesssim \frac{\lambda^{1/8}}{\beta^{1/2}} ~,
\label{Necond}
\eea
where $\lambda$ is the $D3$-charge of the background, e.g.~given by
three-form fluxes.
Namely the field theory results only applies to the short scales
satisfying (\ref{Necond}). Above that, the quantum fluctuations become
stringy. (The zero-mode inflation still proceeds as long as
back-reactions on the background throat, including those from the
relativistic probe brane
\cite{Silverstein:2003hf,Chen:2005ad,Chen:2004hu} 
and those from the de Sitter expansion \cite{Chen:2005ad}, can be 
neglected.)

Although a full understanding of this stringy phase requires a full
string theory treatment, we can see a bound on the scalar fluctuation
amplitude in the following way. In expanding background, the quantum
fluctuations are generated during a Hubble time $\gamma^{-1}
H^{-1}$, and then get stretched out of the horizon and lose causal
contact. Within this period, the brane fluctuates in the transverse
directions, but the fluctuation speed is limited by the local speed of
light $h^2$. So the maximal amplitude for a moving frame observer is
\bea
\Delta r_{max} \approx \gamma^{-1} H^{-1} h^2 ~. 
\label{Drmax}
\eea

To see that it is consistent with our previous arguments, we note that
from field theory, the density perturbation in IR DBI is
\bea
\delta _H \approx H \delta t \approx N_e^2 / \lambda^{1/2} ~,
\label{dHfield}
\eea
and the brane fluctuation is 
\bea
\Delta r = \gamma \delta r \approx \gamma \dot r \delta t 
\approx \gamma h^2 \delta t ~,
\eea
where $\delta t$ is the time delay measured by a lab observer, 
$\dot r$ is the brane speed
(zero-mode scalar speed) which is approximately the speed of light
$h^2$ because the brane travels ultra-relativistically in DBI
inflation. Requiring 
\bea
\Delta r \le \Delta r_{max}
\label{DrDrmax}
\eea
gives the same
condition as (\ref{Necond}).

But the condition (\ref{DrDrmax}) is more transparent in terms of the
scalar fluctuation amplitude. We get
\bea
\delta_H \approx H\delta t \lesssim H \frac{\Delta
r_{max}/\gamma}{h^2} 
\approx 1/\gamma^2 ~.
\label{dHbound}
\eea
(Note that in this formula the time decay $\delta t$ is measured in
the non-moving lab frame. That's why the Lorentz contraction factor
$1/\gamma$ arises in $\Delta r_{max}/\gamma$.)
This means that, although naively from field theory results
(\ref{dHfield}) that $\delta_H$ can grow to one as we look at large
enough $N_e$ , 
it actually has a upper bound
far below one, since $\gamma \gg 1$.
The type of eternal inflation caused by the random walk
of the scalar field is forbidden. 
We note that this type of constraint applies to general DBI inflation 
models.

As we will discuss in the
Sec.~\ref{SecMultithroat}, 
an eternal inflation of false vacuum type, which
is responsible for generating candidate inflaton branes,
generically happens in this scenario.

\section{Possible presence of eternal inflation in brane inflation}
\label{SecEternal}

\subsection{On top of the potential in the bulk}
\label{SecTop}

Because of the compactness of the extra dimensions, 
it is inevitable that some
parts of the inflaton potential profile may have vanishing first
derivative. Especially for those potentials that grow toward the bulk,
such as those we have considered in the KKLMMT slow-roll or UV
DBI inflation, at somewhere in the bulk $\partial V/\partial \phi_{i} =0$
and $\partial^2 V/\partial \phi_{i}\partial \phi_{j}<0$ will appear. 
Classically the inflaton velocity 
can be infinitely small if it stays on the top of such a potential. So
it is interesting to look in more details the conditions for the
random walk type eternal inflation to happen.

Let us simplify the form of such a potential to be $V=V_0 -\half m^2
\phi^2$, where $\phi$ is measured with respect to the top of the potential.
Using the standard slow-roll results, we know that the
density perturbation $\delta_H$ becomes one at the place
\bea
\hat{\phi} = \frac{V_0^{3/2}}{5\sqrt{3} \pi M_p^3 m^2} ~.
\eea
Because of the quantum fluctuations of the inflaton, the precision of
putting a brane on the top of potential is given by $\delta \phi =
H/2\pi$. Therefore we require $\hat{\phi} >H/2\pi$. This is always satisfied
given that $m^2 \ll H^2$, i.e.~the condition that the inflation happens
on the top of such a potential. Therefore eternal inflation of the
random walk type happens in the bulk if such an inflationary potential
can be obtained by fine-tuning, 
for example in the models of
Ref.~\cite{Burgess:2001fx,Iizuka:2004ct}.

It is worthwhile to point out another key difference between brane inflation and 
a more traditional inflationary scenario such as chaotic inflation.
In chaotic inflation, if we are willing to accept a random value for the initial inflaton field 
in the presence of many causally disconnected patches, 
it is reasonable that the inflaton field takes an initial value larger than the 
critical value in at least one patch, thus leading to eternal inflation. 
Even if the inflationary universe was created spontaneously with only one patch, 
one may argue it is more likely that $\phi_{initial} > \phi_{c}$, since $\phi_{initial}$ 
can apriori take an arbitrarily large value in the absence of any selection
principle (such as the wavefunction of the universe). 
In brane inflation, the inflationary universe typically starts
with one or more pairs of $D3-\bar{D}3$ branes, where all values of $\phi$~s are bounded
by the size of the bulk.
First consider the case with only one pair inside one causal patch. 
There is no reason that the $\bar{D}3$ brane starts out sitting precisely at the 
bottom of a throat. In general, it will start somewhere in the bulk and then, due to the 
attractive force, move relatively quickly towards the bottom of the throat. While this is 
happening, its tension will start dropping and the potential for the $D3$ brane will approach the
eventual one that is adopted above. That is, the $D3$ brane 
is seeing a rapidly changing potential as it is moving in the bulk.
While this is happening, both the position and the height of the inflaton potential at 
the top of the hill discussed above have not even emerged yet.  
We expect the $D3$ brane to be dragged towards the throat during this
time.
% (presumably towards the $\bar{D}3$ brane or the throat). 
So, it seems some fine-tuning is needed 
for the $D3$ brane to end up at the top of the hill after the $\bar{D}3$ brane has settled at
the bottom of the throat. Clearly a more careful analysis is warranted.
%It will be interesting to see whether such a scenario actually enhances or suppresses the likeliness of eternal inflation.
Obviously, this pictures gets much more complicated if we start with multi-pairs of $D3-\bar{D}3$ 
branes or some other arrangements. However, in this case 
it seems that the random walk type eternal inflation is possible but 
absent under generic initial conditions. Of course, if inflation comes only from the motion of the D3 brane
(that is, no  $\bar{D}3$ brane), then the random walk type eternal inflation becomes more likely.

We may apply the same argument if the inflaton potential has a local shallow minimum in the bulk.
The attractive force of the $\bar{D}3$ brane plus its motion tend to drag the $D3$ brane towards the throat. Of course, if the minimum is deep enough, then the $D3$ brane may be stuck there. This leads to eternal inflation of the tunneling type.

\subsection{At the tip of the IR throat}

The top-of-a-hill like potential is also possible when we extend the
potential to the tip of a throat, analogous to the potential in the IR
DBI model. The difference is that we are now interested in a
fine-tuned slow-roll shape, $|\beta| \ll 1$.
The same discussion of the last subsection applies except for the
following complication. If the throat is very long, i.e.~extends to
$\phi\to 0$, the slow-roll inflationary phase will smoothly transit to
the DBI inflationary phase at $\phi_{DBI} = \beta \sqrt{N} H$ for
$N>\beta^{-2}$ \cite{Chen:2005ad}. 
This is because towards the IR end of the throat, the
speed-limit constraint becomes more important and have to be taken
into account. 
So if the IR end of the throat is terminated at $\phi
> \phi_{DBI}$, the warping does not change beyond the cutoff, and we
get a random walk eternal inflation near $\hat{\phi}$. 
Otherwise, if the throat is terminated at
$\phi < \phi_{DBI}$, the inflation will become the DBI type before the
slow-roll inflation can reach the critical point of eternal inflation.
In DBI region, the random walk eternal inflation is forbidden as we
saw in Sec.~\ref{SecIR}. However, the warping of a throat cannot
extend to infinity in the IR end, either because of a cutoff or the
back-reaction of the dS space which
will smooth out the warping in 
the IR end of the warped space starting at $\phi
\approx H$ \cite{Chen:2005ad,Chen:2006ni}. Hence the speed limit
constraint will not further increase beyond this point, and the
inflation smoothly changes back to be the slow-roll type 
on the top of the
hill. In this case it is this period of slow-roll inflation that may
be responsible for random walk eternal inflation, if the branes 
start from there.

\subsection{Multi-throat brane inflation}
\label{SecMultithroat}

So far we have concentrated on the eternal inflation of
the Brownian type. We saw that in brane inflation such a
mechanism is either tightly constrained or forbidden for various
different reasons. As mentioned in the introduction, however, eternal
inflation is also possible through a first order phase transition of
false vacuum bubble nucleation. In this and next subsection, 
we discuss this mechanism in brane inflation.

Warped compactification in type IIB string theory is a natural place
to realize the brane inflation. In such a case, there is a very
natural mechanism to generate candidate inflaton branes in the
multi-throat brane inflationary scenario
\cite{Chen:2004gc,Chen:2005ad}.
The AdS throats are sourced by RR and NSNS three-form fluxes localized
at conifold singularities on a Calabi-Yau manifold.
The ${\bar D}3$-branes
will naturally settle down in throats,
since they carry
opposite $D3$ charge to the throats and will be attracted to the IR end.
These antibranes have different lifetimes in different throats
depending on the numbers of fluxes and antibranes, and they 
will eventually annihilate against the fluxes which
source the throats \cite{Kachru:2002gs,DeWolfe:2004qx}. 
If the number $p$ of antibranes is much smaller than the number $M$ of
the RR fluxes, such an annihilation process proceeds through a quantum
tunneling. At the end of this annihilation, $M-p$ number of $D3$ branes
will be created to conserve the total $D3$ charge. 
These $D3$ branes will
either stay or come out of the throats, depending on the sign of the
mass-squared of the moduli potential. In the latter case, they become
a direct source of inflaton for IR DBI inflation; or when they move
into the bulk and re-settle down elsewhere, become an indirect
source of inflaton for
other types of brane inflation such as KKLMMT type slow-roll
inflation or UV DBI inflation. Therefore, in this
multi-throat scenario, the quantum tunneling in
the flux-antibrane annihilation is responsible for creating successful
inflation bubbles. As we know such a bubble creation process is
naturally eternal.

\subsection{Multi-brane inflation}
\label{SecMultibrane}

The next model we want to comment on is the multi-brane inflation
\cite{Cline:2005ty}. This model is motivated to solve the fine-tuning
problems
generic to slow-roll brane inflation models. In most brane inflation
models, flux compactification in Type IIB string theory provides
a robust way to stabilize the dilaton and complex structure moduli of
the Calabi-Yau manifold \cite{Giddings:2001yu}, with addition benefit
of realizing the AdS throats \cite{Klebanov:2000hb}. There are also
non-perturbative effects invoked to stabilize the remaining K\"ahler moduli
\cite{Kachru:2003aw}\cite{deAlwis:2005tf}. But unfortunately, the last
step spoils whatever the flatness of the inflaton potential achieved
by warping. The usual way out is to introduce extra parameters
associated with the corrections to superpotentials and flatten the
potential by fine-tuning. For example the
$\beta$ parameter we've encountered in the KKLMMT scenario must be tuned to 
be small to achieve at least 55 e-folds of inflation
\cite{Kachru:2003sx,Firouzjahi:2005dh}. 

In the multi-brane inflation model \cite{Cline:2005ty}, it is argued
that if the racetrack superpotential is used \cite{Kallosh:2004yh},
the number $N$ of 
mobile branes in the AdS throat can be large  without
destabilizing the K\"ahler modulus. With more than one mobile $D3$
brane in the throat, there's an interesting mechanism to flatten the
potential. The branes start out being confined
in a meta-stable minimum of the potential, and then quantum fluctuation
allows them to tunnel out successively. While this happens,
the meta-stable minimum will get shallower, until all the remaining
branes can roll into the throat all together. The potential can be
very flat when the remaining branes start rolling collectively, and
ensures enough number of e-folds. Thus the potential is flattened
dynamically without any fine-tuning.

Eternal inflation will be quite generic in the
multi-brane scenario. When the branes tunnel out the local minimum
quantum mechanically, the inflaton has to climb uphill, and the
physical volume of the meta-stable vacuum will be expanding
exponentially and creating bubbles of open universe, inside which the branes always
tend to tunnel up and make the bubble de-Sitter space. This is again a
self-reproducing process. Thus in the multi-brane inflationary scenario, the tunneling
effects to flatten the inflaton potential will at the same time
inevitably leads to eternal inflation. However, multi-brane inflation requires some judicious
initial setup, which should be considered as a fine-tuning. With the DBI action, we believe the 
multi-brane inflationary scenario is unnecessary.

\section{Conclusions and implications}

In the paper, we have discussed the possibility of eternal inflation in
different brane inflationary scenarios. We find that eternal inflation
of the random walk type is generically absent in brane inflation.
In the slow-roll scenario, because of the size of the bulk, the critical 
inflaton value for eternal inflation can never be
reached.\footnote{The restriction on the inflaton extension also
restricts the tensor mode in single field slow-roll models as
discussed in Ref.~\cite{Lyth:1996im}.}
For the UV DBI inflation in the
ultra-relativistic regime where the density perturbations does not
run, COBE normalization puts a strong constraint
on the ratio of quantum fluctuation to classical motion. For the IR
DBI inflation
where the density perturbations runs, the speed limit constraint
restricts the amplitude of the scalar quantum fluctuations. These
effects exclude the possibility of random walk eternal inflation in
DBI inflation.

In the case of multi-throat or multi-brane inflation, the generation
of inflaton for 
a successful inflation relies on quantum tunneling process. 
So eternal inflation is
inevitably built into these scenarios. 

Hence, within brane inflation, we see that eternal inflation can be 
inevitable or impossible, depending on the specific scenarios.
Fortunately, these different scenarios of brane inflation have different predictions
on the properties of the cosmic microwave background radiation. More precise 
measurements as well as theoretical analysis in the near future should be able 
to test if brane inflation is correct or not. If the result is positive, it should be able to 
select the particular brane inflationary scenario that our universe has gone through.
In this optimistic case, we will find out if our universe is eternally inflating or not. 

Eternal inflation has lots of implication on quantum
cosmology (see for example \cite{Guth:2000ka}). Eternal inflation generally leads to the 
conclusion that the ultimate initial conditions for the universe become totally
divorced from observation. Because if inflation is eternal into the
future and produce infinite number of Hubble patches, then the
statistical properties of the inflating regions should approach a
steady state independent of initial conditions, even if the underlying
string theory vacuum is not unique. 

The string cosmic landscape contains numerous meta-stable vacua.
We like to find out how nature selects one vacuum versus another vacuum, 
that is, why our particular vacuum is chosen.
There needs to be a mechanism to pick out a particular vacuum
in the string landscape dynamically, like the Hartle-Hawking
wave-function of the universe \cite{Hartle:1983ai} and its improved version 
\cite{Firouzjahi:2004mx,Sarangi:2005cs,Sarangi:2006eb},
or the Hawking-Turok instanton \cite{Hawking:1998bn}. 
Since cosmological data strongly 
suggests that our universe has gone through an inflationary epoch, we have to 
understand how this particular inflationary universe, that is, the one that evolves to 
our today's universe, is selected. We call this SOUP, the Selection of the Original 
Universe Principle \cite{Firouzjahi:2004mx}. 
The question of pre-inflation physics becomes very relevant.  
Ref.~\cite{Firouzjahi:2004mx,Sarangi:2005cs,Sarangi:2006eb} suggest that the 
spontaneous creation of a particular inflationary universe may be selected because 
it has the largest probability of tunneling from nothing (i.e., no classical spacetime).  
%The same mechanism will also be able to distinguish different inflationary scenarios, as the initial condition required by certain models may be harder to realize than others. 

However, if eternal inflation is present in this particular
inflationary universe,
the inflating part of the universe will be exponentially larger than the non-inflating 
part (which includes our observable universe) today. 
This implies that it is exponentially more likely for a random occurrence to take 
place in the part of the universe that is still inflating.{\footnote {There are possible 
alternatives to using volume as the measure, see
e.g. \cite{Garriga:2005av,Bousso:2006ev}.}
This means SOUP will select an inflating part of the universe, and not an inflationary 
universe that has already ended 13.7 billion years ago. That is, SOUP is doomed to fail if 
eternal inflation is present in the particular inflationary universe selected. 
We see in this paper that eternal inflation is not generic in brane inflation. If the improved 
wavefunction selects a brane inflationary scenario that does not lead to eternal inflation, 
then SOUP is still alive as a guiding principle. 
Even in the IR model where eternal inflation may be generic, the success 
of SOUP must be accompanied by a dynamical reason why eternal inflation is absent there.

\acknowledgments
We thank Raphael Bousso, Hassan Firouzjahi, Alan Guth, Shamit Kachru,
Andrei Linde, Liam McAllister, Sarah Shandera and Gary Shiu 
for useful discussions.
XC and SS thank the organizers of the workshop on String Vacua and the
Landscape and the Abdus Salam ICTP for the hospitality.
SHT thanks KITP at UCSB for the hospitality.
The work of XC is supported in part by the Department of Energy.
The work of SS  is supported in part by DOE under DE-FG02-92ER40699.
The work of SHT and JX is supported by the National Science Foundation
under grant PHY-0355005. The research is also supported by the NSF
under grant PHY99-07949.

\appendix

\section{Langevin analysis for DBI inflation}

To study the impact of the quantum fluctuation on the motion of the
inflaton is usually carried out in the
Langevin formalism. One can follow this approach and simply add a
noise term to the inflaton equation of motion. 
For the slow-roll
scenario, this reduces to a rather simple system. For DBI inflation,
a similar procedure can be done 
in the Hamilton-Jacobi formalism and then introduce an
appropriate noise term. 

From the density perturbations of the DBI inflation
\cite{Garriga:1999vw,Alishahiha:2004eh,Chen:2005ad}, the scalar
fluctuations corresponds to a random walk velocity
\bea
\dot \phi_{q} = \frac{H^{3/2}}{2\pi} n(t) ~,
\label{dotphiq}
\eea
where $n(t)$ is defined as in (\ref{n_aver}) and (\ref{n_std}).
Combining (\ref{dotphiq}) with the Hamilton-Jacobi equations
(\ref{eom_V},\ref{eom_gamma},\ref{eom_phidot}), we get the
stochastic equation of motion for DBI inflation
\begin{eqnarray} 
\label{Ldotphi_1}
V(\phi) &=& 3M_p^2H^2(\phi) - T_3h^4(\phi)(\gamma(\phi) - 1) ~, \\
\label{Ldotphi_2}
\gamma(\phi) &=& \sqrt{1 + 4M_p^4T_3^{-1}h^{-4}(\phi)H'^2(\phi)} ~, \\
%\dot{\phi_c} &=& \frac{-2M_p^2H'(\phi)}{\gamma(\phi)} ~, \\
%\dot \phi_{q} &=& \frac{(\gamma H)^{3/2}}{2\pi} n(t') \leq h^2 T_3^{1/2} ~, \\
\dot \phi &=& {\dot \phi_{cl} + \dot \phi_q} ~ \nonumber \\
&=&  \frac{-2M_p^2H'(\phi)}{\gamma(\phi)} + \frac{ H^{3/2}}{2\pi} n(t)
\label{Ldotphi_3}
\end{eqnarray}
Note that the first 2 equations are simply
Eqs.~(\ref{eom_V},\ref{eom_gamma}) unchanged. 
Only Eq.~(\ref{eom_phidot}) is modified with the stochastic noise to
yield Eq.(\ref{Ldotphi_3}). This is a particularly simple way to implement the
noise term into DBI inflation.

It is easy to see how the slow-roll case is reproduced. Recall the
Lorentz factor (\ref{gamma}). In the slow-roll case, $\gamma
\rightarrow 1$, so
\bea
T(\gamma -1) = T\left(\frac{1}{\sqrt{1 - \dot\phi^{2}/T}} -1 \right)
\rightarrow \frac{\dot\phi^{2}}{2}
\eea
which gives the usual kinetic term. Dropping the $\ddot \phi$ in the
inflaton equation of motion reduces Eq.(\ref{Ldotphi_3}) to
\bea
H'(\phi) &=& - \frac{V'}{6H}  \nonumber \\
\dot \phi & =&  -2M_p^2H'(\phi) + \frac{ H^{3/2}}{2\pi} n(t) ~ \nonumber 
\eea
This is the set of equations used in the Langevin
analysis in the slow-roll scenario.

Apriori, the equations of motion do not guarantee that the speed of
the quantum fluctuations satisfy the speed-limit constraint in the
directions transverse to the brane, since the fluctuation
amplitude is determined by the uncertainty principle. But the mean
value of the r.h.s.~of Eq.~(\ref{Ldotphi_3}) stays 
subluminal as long as the quantum fluctuations are within the field
theory regime, namely the Hubble parameter is below the local string
scale.
This is because the speed limit in the transverse brane directions
$\sqrt{T_3} h^2$ restricts the fluctuation amplitude to be less than
$\sqrt{T_3} h^2 H^{-1}$ within a Hubble time. The field theory result
gives $H$, for which to stay subluminal the condition is $H<T_3^{1/4}
h$.
However, we may still have to worry about the Gaussian tail end of the noise, 
which may introduce super-luminal effects. So we must solve the above 
equations in a way that this problem never disrupts its solution. 

The above set of equations may be solved in the following way:  
at an initial $\phi_{j}=\phi (t_{j})$,
suppose we are given $\gamma (\phi_{j})$, so we know both
$T(\phi_{j})$ and $V(\phi_{j})$.
We can use Eq.(\ref{Ldotphi_1}) to determine $H(\phi_{j})$ and use
Eq.(\ref{Ldotphi_2}) to determine $H'(\phi_{j})$. 
Next we treat Eq.(\ref{Ldotphi_3}) as an equation that describes the
time evolution of $\phi$.
Knowing $\phi_{j}=\phi (t_{j})$, this allows us to solve it to find
$\phi_{j+1}= \phi (t_{j} + \Delta  t)$.
We may use Eq.(\ref{Ldotphi_2})
to eliminate $\gamma$ in Eq.(\ref{Ldotphi_1}) so the new
Eq.(\ref{Ldotphi_1}) gives the
$\phi$-evolution of $H(\phi)$. So, knowing $H(\phi_{j})$ and
$H'(\phi_{j})$ allows us to determine 
$H'(\phi_{j+1})$ and so $\gamma (\phi_{j+1})$. 
Repeating these steps allows us to find the Brownian motion of $\phi$
as well as $H(\phi)$.

The actual procedure to solve the above set of equations may be
modified to improve efficiency. 
The key is to treat $\gamma$ as a function of $\phi$ only. This allows
us to treat Eqs.(\ref{Ldotphi_1},\ref{Ldotphi_2}) as constraint equations (for
energy conservation, i.e., the energy density$=3 M_{p}^{2}H^{2}$) and so 
no noise term should be introduced in them. 
So only Eq.(\ref{Ldotphi_3}) contains a noise term.
However, the fluctuation introduced by this noise term appears in $\phi (t)$ 
as well as in $H(t)$ even though $H(\phi)$ remains noise-free. 

The Langevin equation gives the inflaton velocity as an addition of
a drift and a random walk. For slow-roll inflation, both of the
velocities are non-relativistic so the summation is Galilean. For DBI
inflation, at least the former is relativistic, so from this point of
view the velocity addition has to follow the rules of 
the special relativity.
So we can first go to the moving frame on the free-falling 
brane where we only have
simple quantum fluctuations, and then boost it to the lab frame which
is related to the moving frame by a local Lorentz transformation. 
In the following we demonstrate how such a procedure reproduces the
Langevin equation (\ref{Ldotphi_3}).

In the DBI inflation, the brane is moving relativistically with a
slowly changing Lorentz factor $\gamma$.
Let us first look at the quantum fluctuations on the moving frame
which moves instantaneously with the brane for a Hubble time, during
which $\gamma$ remains approximately constant. At each of these
snapshots, the Hubble parameter is $\gamma H$. This is because the
spatial expansion, which is perpendicular to the brane velocity,
proceeds faster by a factor of $\gamma$ due to the special relativity.
Therefore in the mode equation, 
the quantum fluctuations generated and then frozen 
during this period takes the value
\bea
|\delta \phi_k|_{mov} = \frac{\gamma H}{\sqrt{2} k^{3/2}} ~.
\label{dphimov}
\eea
This moving frame is not the best frame to integrate the fluctuations,
since it is only instantaneous. But this is best suited
for us to write down the Langevin equation, which deals with the
instantaneous velocity. 
According to Eq.~(\ref{dphimov}), the variance $\Delta \phi$ generated
during the Hubble time $\Delta t = \gamma^{-1} H^{-1}$ is
\bea
\langle \Delta \phi^2 \rangle_{mov} = 
\frac{1}{4\pi^2} \frac{\Delta k}{k} H^2 \gamma^2 ~.
\label{phi_var_mov}
\eea 
The corresponding instantaneous
velocity of the scalar fluctuations is then
\bea
\dot \phi_{q}|_{mov} = \frac{(\gamma H)^{3/2}}{2\pi} n(t') ~,
\label{dotphiq_mov}
\eea
where $n(t')$ is defined as in (\ref{n_aver}) and (\ref{n_std}), but
now the prime indicates that it is in the moving frame.

As mentioned since the brane is moving relativistically, 
we need to use the relativistic
version of velocity addition,
\bea
\dot \phi 
&=& \frac{\dot \phi_c + \dot \phi_q}{1+ \frac{ \dot \phi_c
\dot \phi_q} {h^4 T_3}} \label{rel_add0} \\
&=& \frac{\dot \phi_c + \frac{(\gamma H)^{3/2}}{2\pi} n(t')}
{1+ \frac{ \dot \phi_c \frac{(\gamma H)^{3/2}}{2\pi} n(t')} {h^4
T_3}}~, \label{rel_add}
\eea
where $\dot \phi_c$ is the classical motion of the brane,
the warped tension $T=h^4 T_3$ is the square of the 
local speed of light which is the same
to both the lab (the throat) and the moving (the $D3$ brane) observers.
In the second step, we work in the field theoretic regime of
the DBI inflation. In this regime,
the brane is moving relativistically while the quantum fluctuations
speed is non-relativistic,\footnote{There is always a small tail in the
distribution of $n(t')$ that violates the speed-limit constraint, which
should be cut off.  But 
for numerical purpose this introduces negligible
corrections as long as the mean value of $\dot \phi_q$ satisfies
(\ref{nonrelq}).}
\bea
\dot \phi_c \approx h^2
T_3^{1/2} ~, ~~~~~
\dot \phi_q \ll h^2 T_3^{1/2} ~.
\label{nonrelq}
\eea 
The quantum fluctuations in the moving frame $\dot \phi_q$ 
is given by
Eq.~(\ref{dotphiq_mov}).

Expand the relation (\ref{rel_add}) we get 
\bea
\Delta \dot \phi =
\dot \phi - \dot \phi_c \approx \dot \phi_q / \gamma^2 ~.
\label{dotphilab}
\eea
This is the relative velocity of the fluctuations to the classical
motion viewed in the lab frame.
In the non-relativistic fluctuation case, the time periods in two
frames are related by
\bea
\Delta t = \gamma \Delta t' (1+ \frac{\dot \phi_c \dot \phi_q}{h^4
T_3})
\approx \gamma \Delta t' ~.
\label{DtDt'}
\eea
Using (\ref{dotphilab}), 
the variance of $\phi$ viewed in the lab frame is
\bea
\langle \Delta \phi^2 \rangle_{lab} 
&=& \int dt_1 dt_2 \langle \Delta \dot \phi_1 \Delta \dot \phi_2
\rangle   \nonumber \\
&=& \Delta t \int \gamma dt'_1 \frac{H^3}{ 4\pi^2 \gamma}
\langle n(t'_1) n(t'_2) \rangle \nonumber \\
&=& \frac{\Delta k}{k} \frac{H^2}{4\pi^2} ~.
\eea
This reproduces the leading order result in the direct calculations
\cite{Garriga:1999vw,Alishahiha:2004eh,Chen:2005ad}. Since this is
viewed in the lab frame, it can be integrated straightforwardly.
This result can be readily understood if we focus on the frozen
fluctuations amplitude \cite{Chen:2005ad}. The amplitude viewed in the
moving frame $\gamma H$ is Lorentz-contracted by a factor of
$\gamma^{-1}$ when viewed in the lab frame, since this amplitude lies
in the same direction as the moving frame velocity. But the horizon
size $(\gamma H)^{-1}$ is perpendicular to the velocity, so it remains
the same to the lab frame.

In fact, in this non-relativistic $\dot \phi_q$ case, 
using (\ref{dotphilab}) and 
(\ref{DtDt'}) one can rewrite the Langevin equation
(\ref{rel_add}) in the following simple way,
\bea
\dot \phi = \dot \phi_c + \frac{H^{3/2}}{2\pi} n(t) ~.
\label{Lv_nonrel}
\eea
Note that the $\gamma$ factors cancel out and the coordinate $t$ in the
stochastic variable $n(t)$ is defined in the lab frame. Under the
condition (\ref{nonrelq}), the mean value of 
the r.h.s.~of this equation is guaranteed
to be smaller than the speed of light. 
We exactly get Eq.~(\ref{Ldotphi_3}).

One can also write down the corresponding 
Fokker-Planck equation for (\ref{Lv_nonrel}) following
Ref.~\cite{Salopek:1990re},
\bea
\frac{\partial P}{\partial t} = \frac{1}{8\pi^2}
\frac{\partial^2}{\partial \phi^2} (H^3 P)
-\frac{\partial}{\partial \phi} ( \dot\phi_c P ) ~.
\eea

In the case where the $\dot \phi_q$ becomes relativistic, we enter a
region of string fluctuations and we have to use
Eq.~(\ref{rel_add0}) to avoid the superluminal fluctuation speed. 
We do not intend to solve the details of this case in this paper, but
we can give a bound on the scalar fluctuation amplitude based on
causality. The scalar
quantum fluctuations speed is now bounded by $\dot \phi_q \leq h^2
T_3^{1/2}$ in the moving frame. 
Correspondingly the fluctuation amplitude is bounded by $\frac{h^2
T_3^{1/2}}{\gamma H}$ in the moving frame and by $\frac{h^2
T_3^{1/2}}{\gamma^2 H}$ in the lab frame. This situation is
discussed in Refs.~\cite{Chen:2005ad,Chen:2005fe} and in
Sec.~\ref{SecIR}.

For some applications \cite{Salopek:1990re}, it seems that the variable 
$T =  \ln(a \gamma H)$ can be useful in the Hamilton-Jacobi setup. One can
straightforwardly rewrite our Langevin equation in term of $T$ as
follows:
\baray
\frac{d\phi}{dT} = - \frac{2M_p^2 H'(\phi)}{\gamma(\phi)}\frac{dt(T)}{dT}
+ n(T), \nonumber \\
<n(T)> = 0, \nonumber \\
<n(T)n(T')> = \frac{H(\phi)^2}{4\pi^2}  \delta(T - T') ~.
\earay

\end{document}